\documentclass[%
 reprint,
 amsmath,amssymb,
 aps,
]{revtex4-2}

\usepackage{graphicx}% Include figure files
\usepackage{dcolumn}% Align table columns on decimal point
\usepackage{bm}% bold math
\usepackage{soul}
\usepackage[colorlinks,linkcolor=blue,citecolor=blue,urlcolor=blue]{hyperref}
\begin{document}

\preprint{APS/123-QED}

\title{High-field immiscibility of electrons belonging to adjacent twinned bismuth crystals}%

\author{Yuhao Ye$^{1}$, Akiyoshi Yamada$^{2,3}$, Yuto Kinoshita$^{3}$, Jinhua Wang$^{1}$, Pan Nie$^{1}$, Liangcai Xu$^{1}$, Huakun Zuo$^{1}$, Masashi Tokunaga$^{3}$, Neil Harrison$^{4}$, Ross D. McDonald$^{4}$, Alexey V. Suslov$^{5}$, Arzhang Ardavan$^{6}$, Moon-Sun Nam$^{6}$,  David LeBoeuf$^{7}$, Cyril Proust$^{7}$,  Beno\^{\i}t Fauqu\'{e}$^{8}$, Yuki Fuseya$^{2}$, Zengwei Zhu$^{1,}$}
\email{zengwei.zhu@hust.edu.cn}
\author{Kamran Behnia$^{9}$ }

\affiliation{(1) Wuhan National High Magnetic Field Center and School of Physics, Huazhong University of Science and Technology, Wuhan 430074, China\\
(2) Department of Engineering Science, University of Electro-Communications, Chofu, Tokyo 182-8585, Japan\\
(3)Institute for Solid State Physics, The University of Tokyo, Kashiwa, Chiba 277-8581, Japan\\
(4) MS-E536, NHMFL, Los Alamos National Laboratory,Los Alamos, New Mexico 87545, USA\\
(5)National High Magnetic Field Laboratory, 1800 E. Paul Dirac Drive, Tallahassee, FL 32310, USA\\
(6) Department of Physics, Clarendon Laboratory, University of Oxford, Oxford OX1 3PU, United Kingdom\\
(7) Laboratoire National des Champs Magn\'{e}tiques Intenses (LNCMI-EMFL), CNRS, UGA, UPS, INSA,
Grenoble/Toulouse, France\\
(8) JEIP,  USR 3573 CNRS, Coll\`{e}ge de France, PSL Research University, 11, place Marcelin Berthelot, 75231 Paris Cedex 05, France\\
(9) Laboratoire Physique et Etude de Mat\'{e}riaux (CNRS-UPMC)\\
ESPCI Paris, PSL Research University, 75005 Paris, France\\
}

\date{\today}

\begin{abstract}
Bulk bismuth has a complex Landau spectrum. The small effective masses and the large g-factors are anisotropic. The chemical potential drifts at high magnetic fields. Moreover, twin boundaries further complexify the interpretation of the data by producing extra anomalies in the extreme quantum limit. Here, we present a study of angle dependence of magnetoresistance up to 65 T in bismuth complemented with Nernst, ultrasound, and magneto-optic data.  All observed anomalies can be explained in a single-particle picture of a sample consisting of two twinned crystals tilted by 108$^{\circ}$ and with two adjacent crystals keeping their own chemical potentials despite a shift between chemical potentials as large as 68 meV at 65 T. This implies an energy barrier between adjacent twinned crystals reminiscent of a metal- semiconductor Schottky barrier or a p-n junction. We argue that this barrier is built by accumulating charge carriers of opposite signs across a twin boundary.
\end{abstract}

\maketitle

\begin{figure*}
\includegraphics[width=16.6cm]{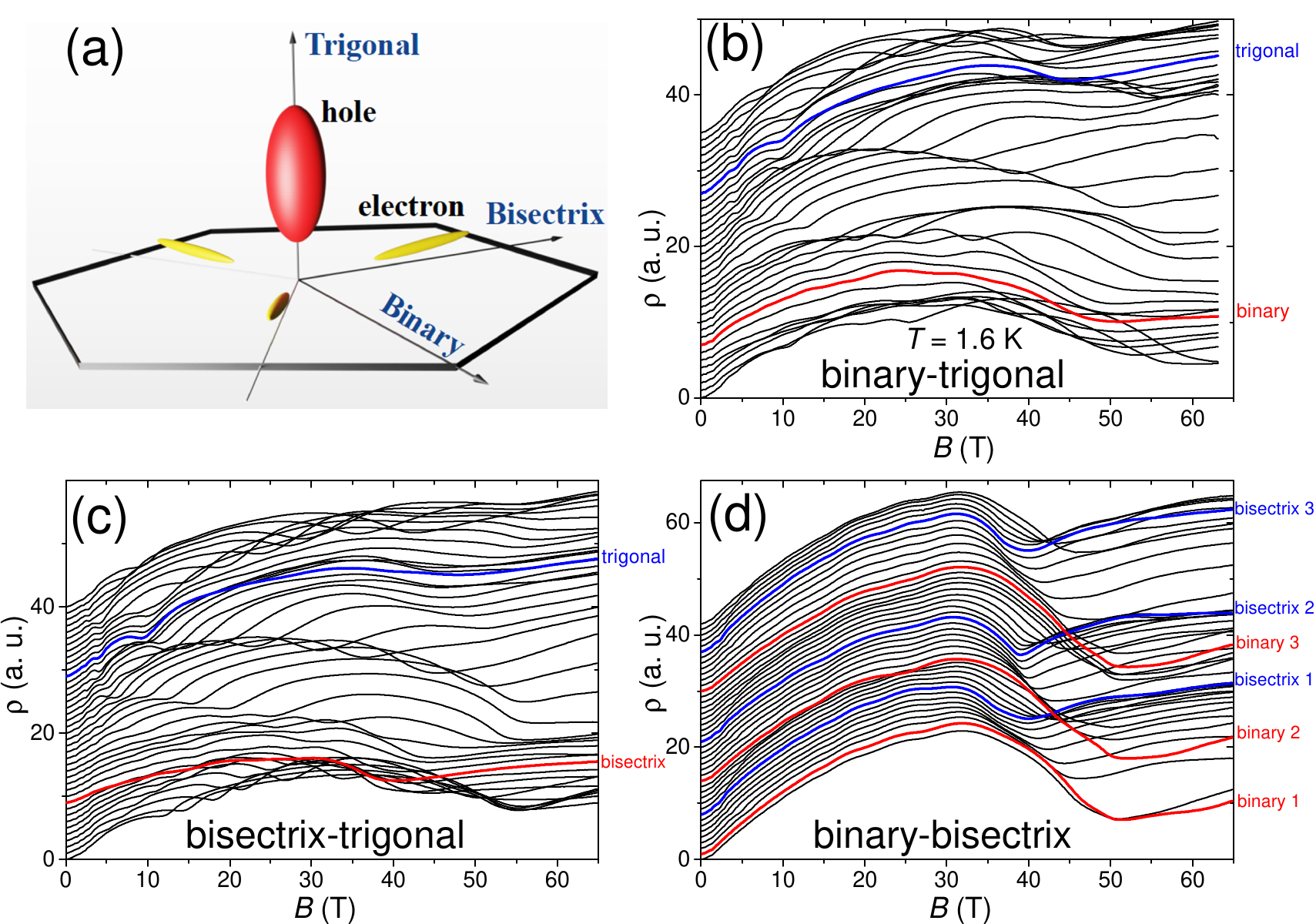}
\caption{\textbf{Magnetoresistance data in the three planes of rotation.} \textbf{a} Sketch of Fermi surface of bismuth, containing a hole and three cigarette-like electron pockets. Magnetoresistance up to 65 T at 1.6 K every $\sim$ 5 degrees in the (\textbf{b}) binary-trigonal, (\textbf{c}) bisectrix-trigonal, and (\textbf{d}) binary-bisectrix planes. The quantum limit of holes, when the penultimate hole Landau level is evacuated, occurs at 9 T when the magnetic field is along the trigonal axis and shifts upward as the field is tilted towards the binary or bisectrix axis.}
\label{Fig:Rrawdata}
\end{figure*}
\setlength{\parindent}{0pt}{\textbf{INTRODUCTION}}

Bismuth occupies a distinct place in the history of solid state physics (For reviews, see \cite{Dresselhaus1971,Edel'man1977,Issi1979,Fuseya2014}). Its tiny Fermi surface was the first to be experimentally mapped out thanks to the easily detectable quantum oscillations it generates \cite{Dhillon1955,Shoenberg1939,Smith1964,Bhargava1967,Liu1995,Kumar2023}. Recent attention focuses on finding topologically non-trivial states both for the surfaces and the hinges of bismuth crystals \cite{Fu2007,Drozdov2014,Benia2015,Aguilera2015,Ito2016,Murani2017,Fuseya2018,Schindler2018,Nayak2019,Hsu2019}. %, postulated to reside on the hinges of bismuth crystals \cite{Schindler2018}. 
Recent transport studies on macroscopic crystals have revealed  a non-trivial variation of the magnetoresistance \cite{Collaudin2015} and the Seebeck effect \cite{Spathelf2022} with the amplitude and the orientation of the magnetic field. A robust boundary conductance emerges at quantizing magnetic field, which is yet to be understood \cite{Kang2022}. At the atomic scale, a bismuth layer on a NbSe$_2$ substrate was found to be a platform for detecting nanometric Turing patterns  \cite{YukiTuring}.

The rhombohedral structure of crystalline bismuth at ambient pressure is the outcome of lowering the cubic symmetry by pulling the cube along its body diagonal axis. Since this can be achieved along each of the four body diagonals, twin formation is common in bismuth crystals \cite{Savenko1998,Bashmakov2002} and twin boundaries in bismuth crystals were directly observed decades ago with a scanning tunneling microscope \cite{Edelman1996,Edelman2005}. 

The uncontrolled presence of such twin boundaries misled researchers seeking the signatures of collective electronic effects in bismuth. In the extreme quantum limit, when all carriers are confined to their lower Landau levels, a variety of electronic instabilities are conceivable  \cite{Halperin_1987,Macdonald_1987}. Experiments seeking such instabilities \cite{Behnia2007c,Li2008} explored bismuth beyond its quantum limit, achievable by applying a magnetic field exceeding 9 T along the trigonal axis, and found anomalies which, at first sight, could not be explained in the single-particle picture. Taken at face value, they were attributed to instabilities caused by electronic interactions. However, these interpretations in the case of bismuth were rejected subsequently. The computed Landau spectrum \cite{Alicea2009,Sharlai2009,Zhu2011,Zhu2012a} was found to be in agreement with the experimental data and the hysteresis attributed to a first-order phase transition \cite{Li2008} was not reproduced by torque magnetometry \cite{Fauque2009b} or magnetostriction \cite{Kuchler2014} measurements.

\begin{figure*}
\includegraphics[width=18cm]{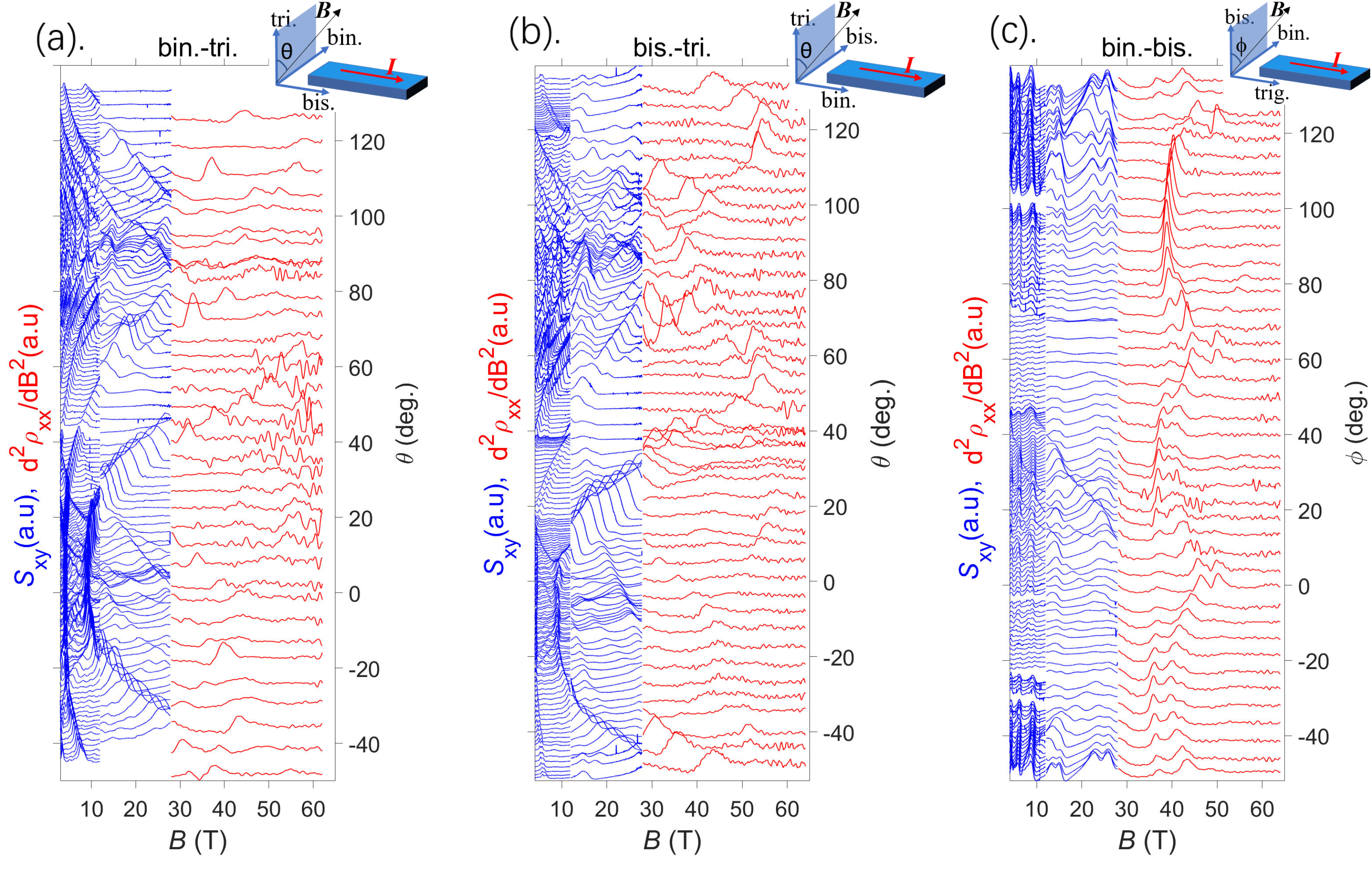}
\caption{\textbf{Field dependence of the second derivative of magnetoresistance and the Nernst response in three rotating planes.} \textbf{a}-\textbf{c} Field dependence of the Nernst voltage ($S_{xy}$) $vs$ magnetic field measured in a magnetic field up to 12 T (in a superconducting magnet), up to 28 T in resistive water-cooled magnet (curves in blue) along with the second derivative of the magnetoresistance measured in pulsed field experiments above 28 T up to 65 T (curves in red) for the three planes: binary-trigonal, bisectrix-trigonal, and binary-bisectrix planes, respectively. The data have been normalized to their highest amplitude for each data set. The 0$^{\circ}$ and 90$^{\circ}$ represent the field along trigonal and binary/bisectrix for the (\textbf{a})/(\textbf{b}). The 0$^{\circ}$ and 90$^{\circ}$ represent the field along binary and bisectrix for the (\textbf{c}). The sketches show the direction of the orientation of the magnetic field and electrical/heat current for each plane of rotation. Note that in all the experiments, the electrical or the heat current is along the direction perpendicular to the plane of rotation.}
\label{fig:raw-data}
\end{figure*}

It was later found that the additional anomalies \cite{Behnia2007c} were caused by the presence of a secondary twinned crystal tilted by 108$^{\circ}$ with respect to the main crystal \cite{Zhu2012a}. Thus, up to 28 T, the single-particle picture proved to be sufficient to explain the experimental data \cite{Zhu2012a}. The same theoretical frame was then employed to explain how the evacuation of one or two electron pockets by a 40 T magnetic field, applied along the binary or the bisectrix axes, generates a sudden drop in magnetoresistance \cite{Zhu2017} and an anomaly in magnetization \cite{Iwasa2019}. 

Theory has predicted that  when carriers are confined to their lowest Landau levels,  the three-dimensional electron gas can suffer a variety of instabilities  \cite{Halperin_1987,Macdonald_1987}. The success of the single-particle picture in bismuth pushed beyond the quantum limit is to be contrasted with the case of graphite, another low-carrier semi-metal, in which the quantum limit is accessible in a three-dimensional context. A succession of field-induced phase transitions have been detected in graphite\cite{Yaguchi_2009,Fauque2013,Wang2020,Zhu2019} indicating the capacity of strong Coulomb interaction to generate collective states \cite{Halperin_1987,Macdonald_1987} in this material. The two elemental semi-metals differ in several aspects. Screening is weaker in graphite because of its lower dielectric constant. Spin-orbit coupling is much stronger in bismuth. Finally, graphite is much more anisotropic than bismuth, and its electron and hole Fermi pockets' longer axis are aligned. 

Three-dimensional semimetals have been found to display numerous interesting properties in presence of strong magnetic field. Let us cite two examples. In ZrTe$_5$, for example, there is a magnetic freeze-out beyond the quantum limit and spin-polarised electrons produce an anomalous Hall effect \cite{Gourgout2022}. In micro-structured Cd$_3$As$_2$, current beams can be guided by tilting a magnetic field \cite{Huang2020}.

\begin{figure*}
\includegraphics[width=17cm]{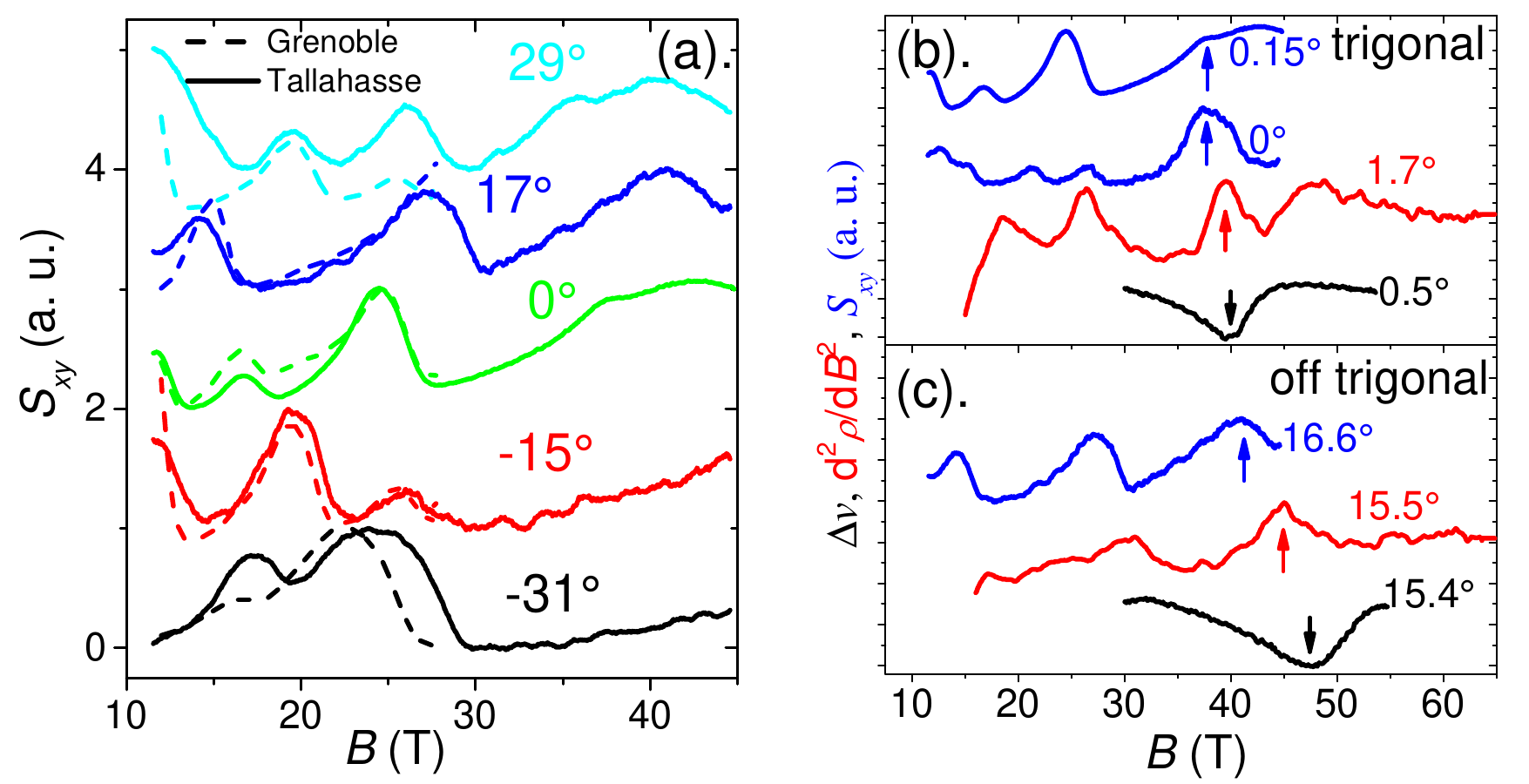}
\caption{\textbf{Consistency among different probes and samples. }\textbf{a} Nernst voltage ($S_{xy}$) measured in Grenoble up to 28 T (dashed lines) compared with the Nernst voltage measured up to 45 T (solid lines) for several orientations of the magnetic field. \textbf{b} Nernst voltage (blue) compared with the second derivative of the magnetoresistance  (red), and the sound velocity  (black) {\it{vs.}} magnetic field at 0$^{\circ}$  (\textbf{c}) Same for   15$^{\circ}$. Peaks in $S_{xy}$, and $\frac{d^2\rho}{dB^2}$ are concomitant with a minimum in $\Delta v$. The arrows indicate the $\sim$40 T anomaly observed previously \cite{Fauque2009}. The measurements presented in all panels were performed in the bisectrix-trigonal plane and zero degree corresponds to field orientation along the trigonal axis.}
\label{fig:techniques}
\end{figure*}

In the case of bismuth-antimony alloys, possible signatures of field-induced  electronic instability were recently observed  \cite{Kinoshita2023}. They are to be distinguished from high-field anomalies due to the presence of twin boundaries seen before \cite{Izawa2008}. 

Here, we present an extensive study of the Shubnikov-de Haas effect, Nernst effect, ultrasound, and magneto-optics in bismuth at low temperatures and a magnetic field as strong as 65 T. We map the angle dependence of the Landau spectra for the whole solid angle and find that all anomalies can be explained in the single-particle picture, invoking the presence of a secondary twinned crystal. This includes the anomaly observed around 38 T when the magnetic field is applied along the trigonal axis of the main crystal \cite{Fauque2009}, whose origin has not been hitherto identified. Thus, up to 65 T, there is no signature of instability caused by electron-electron interaction in bismuth. However, the quantitative agreement between theory and experiment is obtained thanks to a crucial assumption: that the two adjacent twinned crystals each have their own chemical potential $\mu$, set by the orientation of the magnetic field respective to their crystalline axes. We will show that without this assumption, there is no agreement between the computed and the measured Landau spectrum. This unavoidable assumption leads us to conclude that electrons residing in two adjacent crystals do not mix up, despite a 68 meV difference in their chemical potentials at 65 T. We argue that this hypothesis points to the existence of a type of p-n junction at the twin boundary of this semi-metal.\\ 

\begin{figure*}
\includegraphics[width=16.8cm]{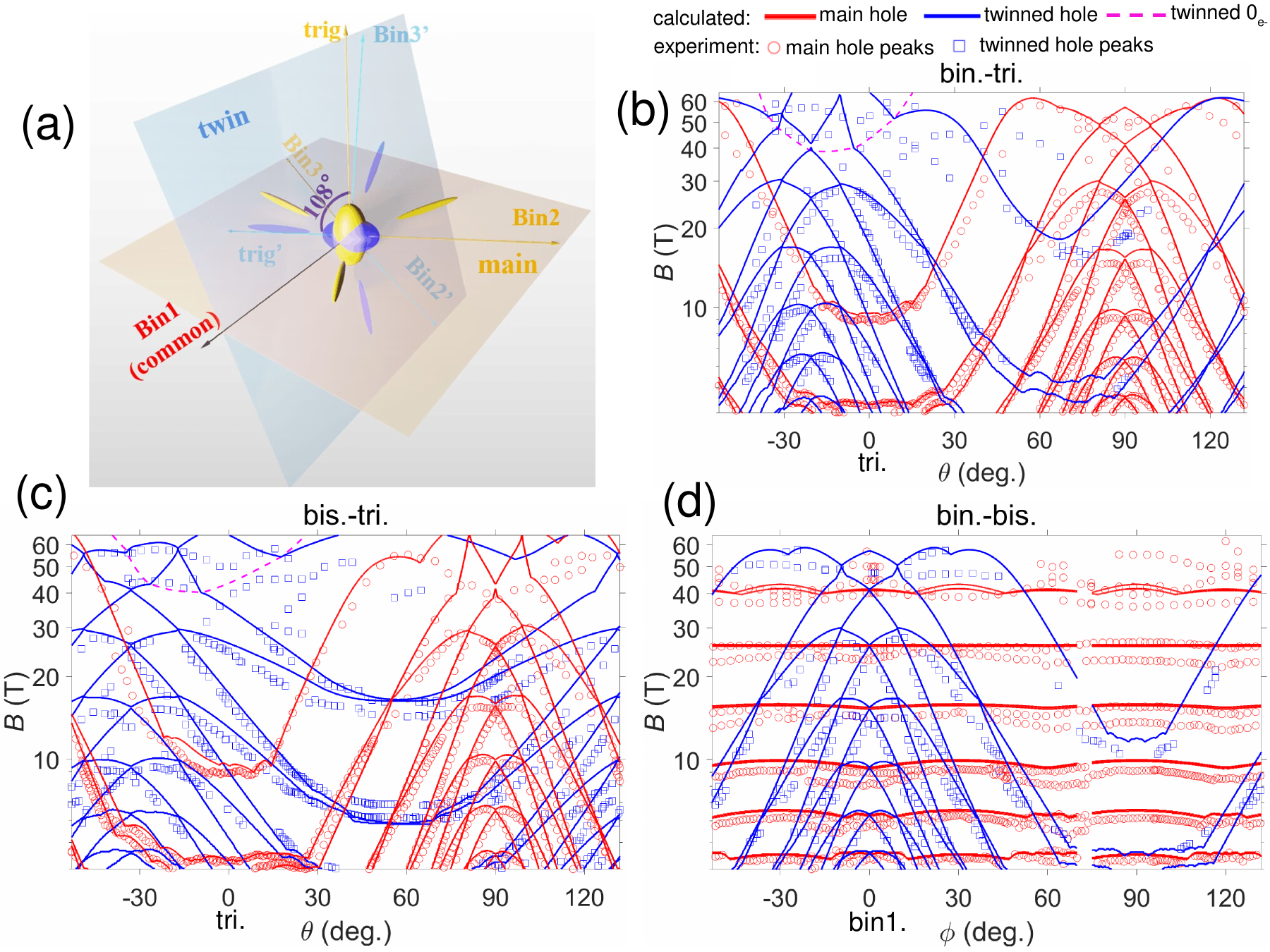}
\caption{\textbf{The experimental and theoretical Landau spectrum. } \textbf{a} The illustration of the twinned structure in bismuth. The two crystals tilt from each other by 108$^{\circ}$ and share one common binary. \textbf{b} The Landau spectrum of the main and twinned samples from holes according to theory (solid lines) and experiment (symbols) for binary-trigonal, (\textbf{c}) bisectrix-trigonal, and (\textbf{d}) binary-bisectrix planes. The red and blue symbols are for the peaks from the holes of the main and twinned crystals from the Fig. \ref{fig:raw-data}, respectively. The red and blue lines represent the calculated Landau levels from holes of the main and twinned domains, respectively. The magenta dashed lines are for calculated 0$_{e-}$ electron Landau level which corresponds to the large drop in magnetoresistance as the field is along trigonal. The other electron peaks below 28 T are omitted for clarity, see the ref \cite{Zhu2012a}.}
\label{fig:peaks}
\end{figure*}

\textbf{RESULTS}\\
\textbf{Angle-dependent magnetoresistance in three planes}

Fig.\ref{Fig:Rrawdata}(a) shows the sketch of the Fermi surface of bismuth, which contains a hole pocket at $T$ point and three cigarette-like electron pockets at $L$ points. Fig.\ref{Fig:Rrawdata}(b)-(d) shows magnetoresistance along three planes, namely binary-trigonal (bin.-tri.), bisectrix-trigonal (bis.-tri.), and binary-bisectrix (bin.-bis.) at 1.6 K. Magnetoresistance shows oscillations which correspond to evacuations of the Landau levels with magnetic field. The different orientations for these planes are indicated. When the magnetic field is along the trigonal axis, MR shows a minimum around 9 T, which corresponds to the evacuation of the penultimate Landau level of hole-like carriers \cite{Bompadre2001}. As the magnetic field is tilted off the trigonal axis, the cross-section of the hole Fermi pocket perpendicular to the magnetic field increases and this minimum shifts to higher fields. In the binary-bisectrix of Fig.\ref{Fig:Rrawdata}(d), drops around 40 T are due to the emptying of Dirac pockets\cite{Zhu2017}.

Fig. \ref{fig:raw-data} presents the second derivative of our magnetoresistance next to the Nernst measurements ($S_{xy}$) for three high-symmetric planes \cite{Zhu2012a}. The two types of data are consistent with each other, but the sensitivity of the Nernst results is relatively higher. Each peak in the Nernst response or in the second derivative of magnetoresistance corresponds to an intersection between an electron (or a hole) Landau level and the chemical potential. Note that the data have been normalized to the largest amplitude of each data set. Fig.\ref{fig:raw-data} (a), (b), and (c) show the data for the magnetic field rotating in the binary-trigonal, bisectrix-trigonal, and in the binary-bisectrix planes, respectively. In the corner of the panel, the schematic diagram shows the measurement configurations for MR. The quantum oscillations pattern is relatively simple and easily understood when the field is along the trigonal axis \cite{Behnia2007bismuth}. The hole Landau peaks can be easily identified as their amplitudes are the largest, except for a sudden jump around 40 T along the trigonal axis.\\

\textbf{Consistency of measurements}

We verified the consistency between diverse techniques when the field was rotated in the bisectrix-trigonal plane and the data obtained on different samples. Fig.\ref{fig:techniques}(a) shows a comparison of the Nernst data collected in Grenoble up to 28 T (dashed lines) with the data collected in Tallahassee up to 45 T (solid lines) for several orientations of the magnetic field. The Nernst data shows an anomaly at 40 T, as reported previously \cite{Fauque2009}. In various experiments, the peak positions are close to each other. Panels (b) and (c) compare three distinct sets of data obtained in different laboratories at different times and on different samples. The Nernst data, S$_{xy}$, the second derivative of magnetoresistance, $\frac{d^2\rho}{dB^2}$, and the change in the sound velocity ($\Delta v$), as the ultrasound propagated along the trigonal direction, are shown for two orientations of the magnetic field, when it is oriented almost along the trigonal axis (b) and when it is 15$^\circ$ (c) off the trigonal axis. There is a good consistency between the three experimental probes, indicating similar peak positions. Specifically, there are $\sim$40 T  anomalies in $S_{xy}$, in $\frac{d^2\rho}{dB^2}$, and in  $\Delta v$. The slight difference in positions at $\theta\approx 15^\circ$ is consistent with the experimental margin in measuring the angle between the magnetic field and the sample in three different experiments. We also notice that the anomalies in sound velocity and ultrasound attenuation occur at slightly different magnetic fields\cite{Akiba2018}.\\

%which is consistently observed across the three experimental probes. 

%Fig. \ref{fig:techniques}(b) and (c), we present the field-dependent Nernst signals from Tallahasse, incorporating data from alternative techniques, such as the second derivative of magnetoresistance and ultrasound, acquired for fields aligned both along the trigonal axis and at 15$^\circ$. Strikingly, all three techniques demonstrate analogous field positions as the Landau levels intersect the chemical potential. The anomaly at 40 T\cite{Fauque2009} is marked by arrows, which is consistently observed across the three experimental probes. It's important to mention that a slight variance in field position at $\theta\approx 15^\circ$ arises from the subtle variations in measurement angles employed by different techniques.

%We extended the Nernst measurement to 45 T using a hybrid magnet in Tallahassee, as depicted in Fig.\ref{fig:techniques}(a) represented by the solid lines, while the dashed lines, detailed in the reference \cite{Zhu2012a}, correspond to the Nernst data obtained from Grenoble at several distinct angles. These lines exhibit quite good consistency. 
\begin{figure}
\includegraphics[width=8.5cm]{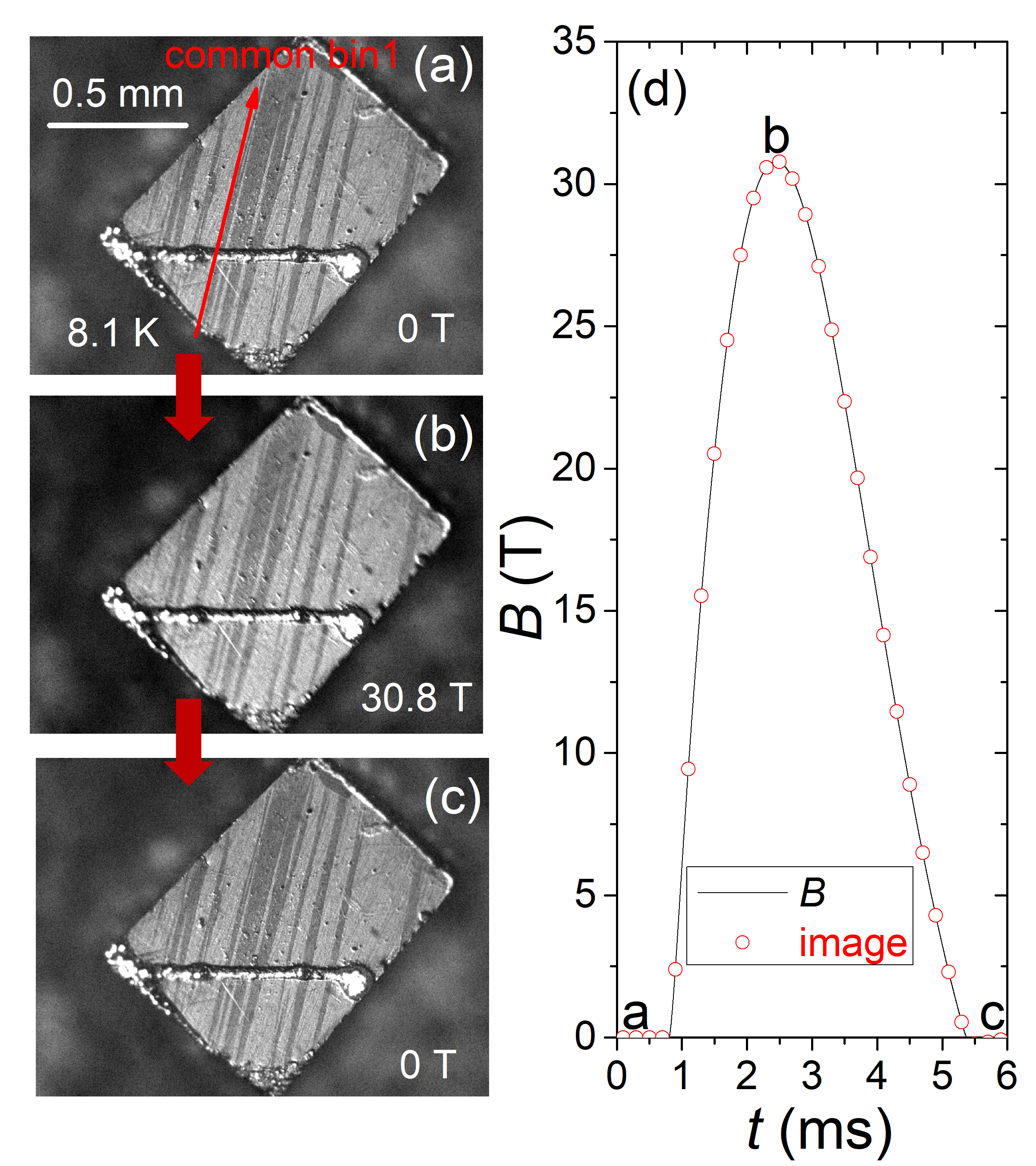}
\caption{\textbf{No detectable change of twins under pulsed magnetic field.} \textbf{a}, \textbf{b}, and \textbf{c} show photos at the specific time at 0 T, 30.8 T, and 0 T after the pulse, indicated in the (\textbf{d}). The crystalline orientations are marked in the (\textbf{a}). \textbf{d} The time pattern of the magneto-optical measurements performed at 8.1 K.  The magnetic field pulse for the imaging experiment lasts $\sim$ 5 ms. Images were taken every 0.2 ms as marked by red open circles. }
\label{fig:MO}
\end{figure}

\begin{figure*}
\includegraphics[width=14cm]{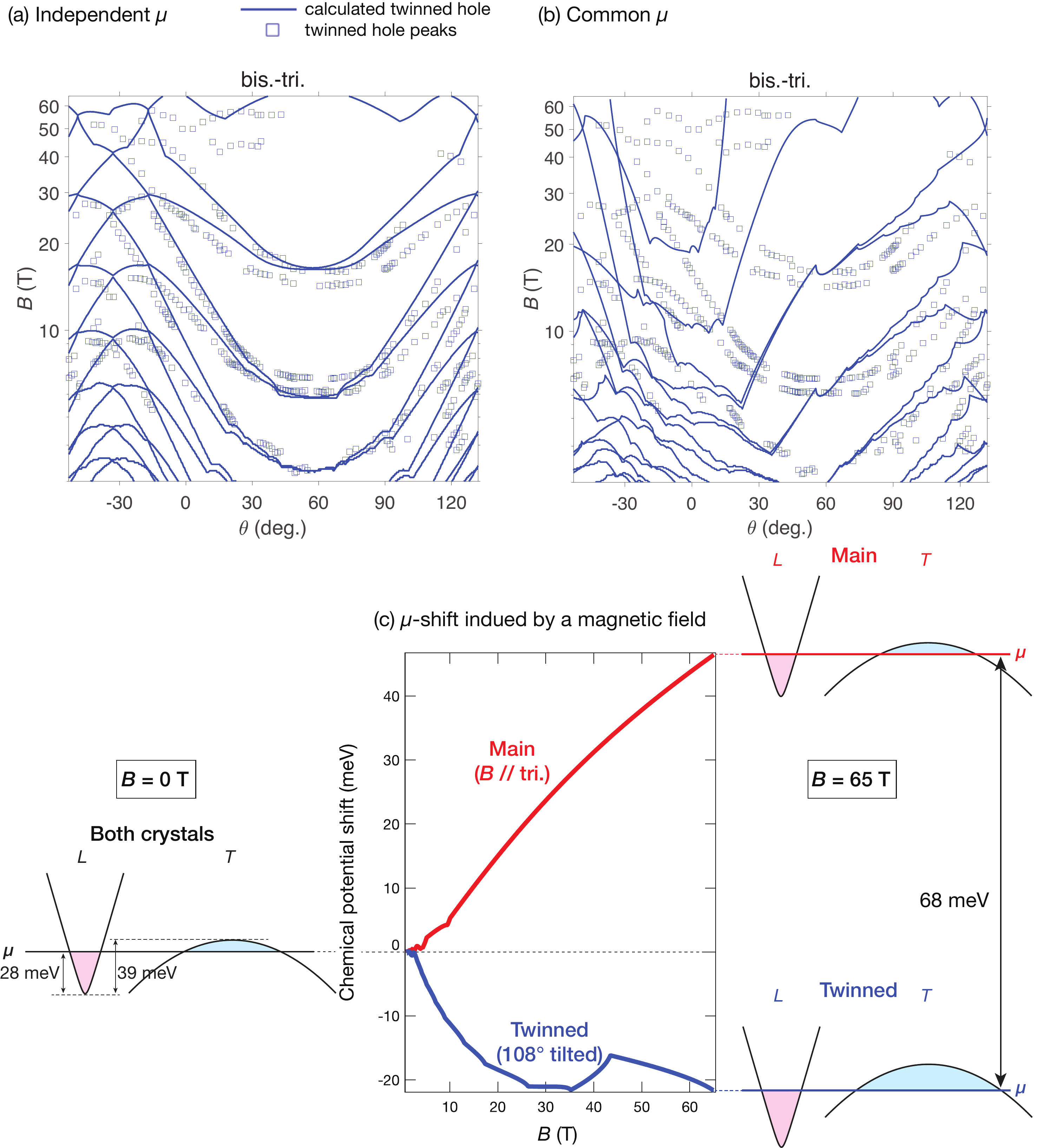}
\caption{\textbf{Comparison between theoretical and experimental results for  independent and common chemical potential.}  The Landau spectrum in the bisectrix-trigonal plane. In (\textbf{a}), $\mu$ in each crystal is independent of the other. In (\textbf{b}), the two crystals share a common $\mu$. The calculated Landau spectrum and the experimental results agree in (\textbf{a}), but radically differ in (\textbf{b}). \textbf{c} The chemical potential shift for the main and twinned sample according to the theoretical results with independent $\mu$. The difference becomes as large as 68 meV at 65 T.}
\label{fig:EF}
\end{figure*}

\begin{figure*}
\includegraphics[width=15cm]{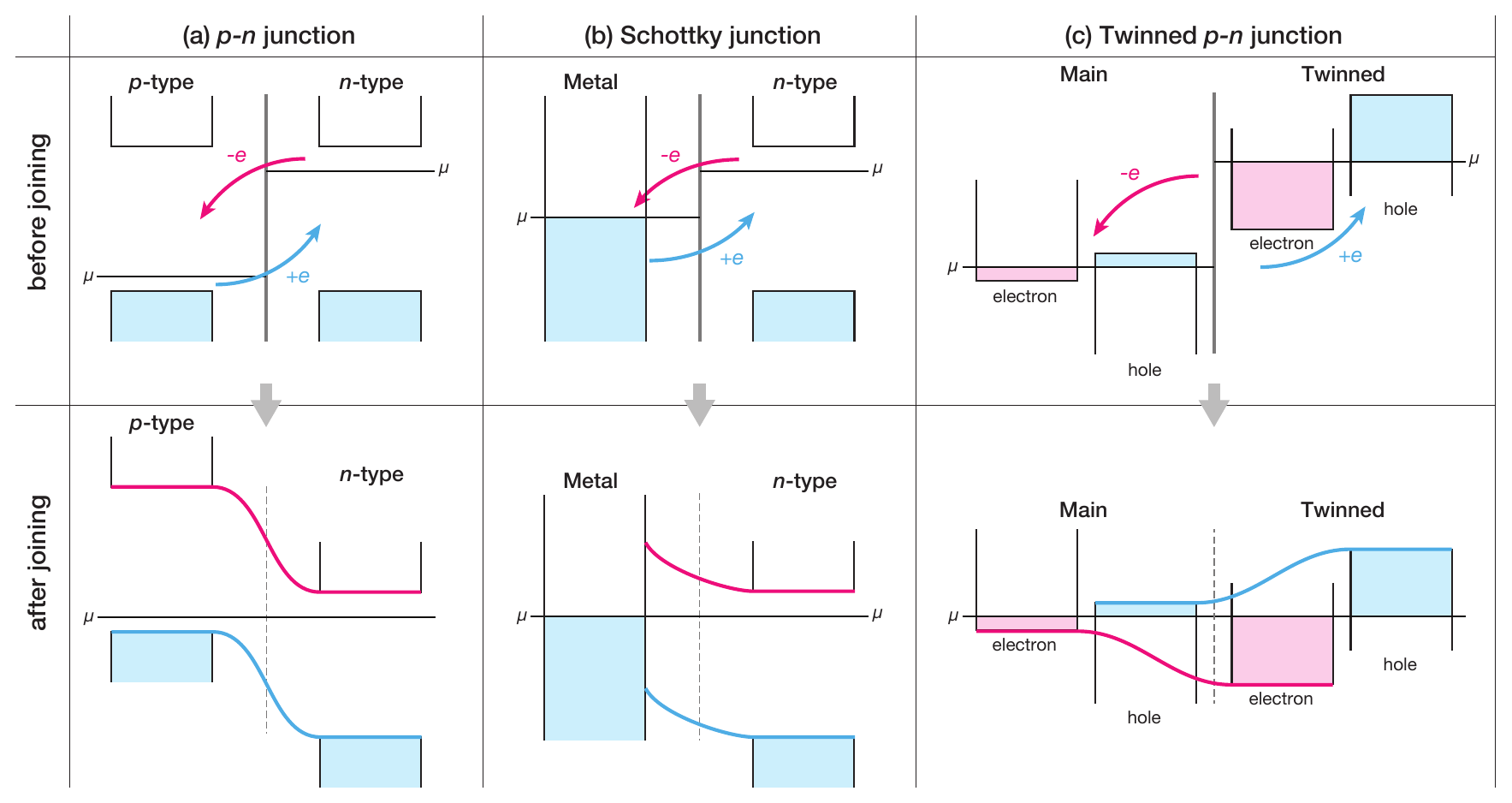}
\caption{\textbf{Schematic of band bending in p-n junctions, Schottky junction and proposed twinned p-n junction in semimetal bismuth before and after joining.}  Band bending in (\textbf{a}) the p-n junctions between two semiconductors; (\textbf{b}) the Schottky junction between a metal and a n-type semiconductor; (\textbf{c}) the twin boundary in semimetal bismuth }
\label{Fig_boundary}
\end{figure*}

\textbf{Comparison with theory}

The angle-dependent Landau spectrum can be described by a theoretical model that considers electron pockets at the L points of the Brillouin zone in an extended Dirac Hamiltonian and the hole pocket at the T point as carriers with parabolic dispersion and highly anisotropic g-factor\cite{Zhu2017,Zhu2011review,Zhu2012a,Fuseya2015}. In order to explain additional and unexpected anomalies, one needs to assume the presence of an additional secondary twinned domain \cite{Zhu2012a,Zhu2017}. In the high field limit, when only a single Landau level is occupied, the interband coupling becomes important and plays a significant role in setting the details of the spectrum \cite{Zhu2017}. 

The twinned sample originates from the distortion by the symmetry-lowering operation due to Peierls transition \cite{Peierls1991}. Since the Schmid factor is 0.48 in bismuth \cite{Ostrikov1999}, applying an external load \cite{Bashmakov2002} and thermal cycling \cite{Ostrikov2001} could relatively easily induce twins compared to other materials \cite{Ostrikov1999}. Thus, the strain appearing during sample cleaving by a blaze along cleavage basal trigonal (0001) plane in liquid nitrogen could introduce twinned domains on the sample and these twinned domains should mainly be on the surface. This could be a reason that a smaller RRR was found in a sample with a smaller dimension \cite{Galev1981}, since the ratio of the one plane area to the volume is smaller in a smaller sample.  When the sample is twinned, besides the existing principal trigonal, a quasi-trigonal axis will be present. 

Fig.\ref{fig:peaks} compares the theoretical prediction with experimental peaks for three high symmetrical planes. The experimental peaks depicted by open symbols are compared with the theoretical results, which are presented by lines from both the main and a secondary tilted crystal as a function of log($B$) (See the Supplementary Figure 2 for a linear plot). For clarity, we have omitted the electron peaks below 28 T, which have been identified earlier \cite{Zhu2012a}, By considering all contributions from both electrons and holes, as well as the main and twin crystal crystals, the agreement between experiment and theory can be further improved (See Supplementary Figure 3 for all data sets). Beyond 28 T, we included all peaks associated with electrons from both domains. Fig.\ref{fig:peaks}(a) sketches the main and twinned crystals with their Fermi surfaces. The tilt angle between the two crystals is 108{$^\circ$}\cite{Zhu2012a}. The common binary axis is also shown. All theoretical results of the hole and twinned electron Landau level 0$_{e-}$ are included. 

The theoretical model here is fundamentally the same as the one used before \cite{Zhu2011,Zhu2012a,Zhu_2018}, except that the tensor, which is the correction to the additional g-factor in the lowest Landau level $V'$, was adjusted to fit the experimental Landau spectra better. Equations (14), (22) from the Ref. \cite{Zhu2011} and (37), (38) from the Ref. \cite{Zhu_2018} were used for calculations. $V_3'$ was adjusted to -0.0025 from -0.0625 \cite{Zhu2017} (See Supplementary Note 4 for a comparison). 
 As seen in Fig.\ref{fig:peaks}, whereas a slight displacement occurs, which may be due to misalignment during measurement, there is a relatively good agreement between the theory and the experiment. Some more peaks may be due to the electron Landau levels, which have not been clearly identified in the figures, see Fig.\ref{fig:peaks}(c). We may conclude, therefore, that the Landau spectrum of bismuth corresponds to what is expected in the single-particle theory. No collective effect due to electron-electron interaction is visible up to 65 T. Because of extreme sensitivity, nullification of the Zeeman splitting of holes along the trigonal axis, the experimentally resolved hole peaks are split due to a small residual misalignment in the case of binary-bisectrix plane (see Fig.\ref{fig:peaks}(d)).  The twinned crystal shares one common binary axis with the original crystal, but tilts its trigonal axis by 108$^{\circ}$ off\cite{Zhu2017}. This is demonstrated by Fig.\ref{fig:peaks}(d) at zero degrees where two sets of hole peaks merge. The additional peaks vanish as the two crystals are in the same configuration regarding the magnetic field and possess an identical Landau spectrum.

The twinning scenario provides an explanation for the previously reported drop in the magnetoresistance for a magnetic field of 38 T along the trigonal axis of the main crystal \cite{Fauque2009}. We can identify it as the result of emptying the electron pockets of the twinned domain. As seen in Fig.\ref{fig:peaks}(b) and Fig.\ref{fig:peaks}(c), the magenta dashed line, which represents the intersection between the 0$_{e-}$ Landau level of the twinned domain and the chemical potential happens to be at $\sim$ 40 T when the magnetic field is along the trigonal axis of the main crystal. Thus, there is no ultra-quantum feature in the main crystal beyond the quantum limit of holes, which occurs at 9 T along the trigonal axis.\\

\textbf{Twins observed by magneto-optics}

We confirmed the presence of twin boundaries in our bismuth samples with a magneto-optical probe and found that pulsed magnetic fields have no influence on the domain population. The \textit{in-situ} images are shown in  Fig.\ref{fig:MO}. The crystal was illuminated with polarized light and the image was recorded using a high-speed camera in pulsed magnetic fields up to 30.8 T oriented along the trigonal axis. The field profile is shown in  Fig.\ref{fig:MO}(d). Images are taken at the magnetic field at the times represented by red circles. Three typical images taken at 0 T, 30.8 T, and 0 T after the pulse are shown in  Fig.\ref{fig:MO} (a), (b), and (c), respectively.  The binary axis common to the two domains is labelled in Fig.\ref{fig:MO}(a). The twinned domains occupy a large area close to one-third of the surface (Please see Supplementary Note 1 for more similar results on the observation of twin domains in another sample.). \\

\textbf{Independent or common chemical potentials for main and twinned crystals}

Let us now turn our attention to the most striking aspect of our results. Since the charge is conserved, any change in the density of electrons is to be compensated by an identical change in the density of holes. This leads to a shift in the position of the chemical potential with increasing magnetic field \cite{Smith1964,Zhu2011,Zhu2012a,Zhu2017}. This field-induced shift strongly depends on the orientation of the applied magnetic field. The main crystal and its twinned domain are tilted by 108 degrees. Therefore, the magnitude of the field-induced chemical potential shift is different for the two twinned crystals. Thus, the Landau spectrum would be very different in two imaginable cases: a) The two crystals keep their own distinct chemical potential; b) The two crystals share the same chemical potential. Fig.\ref{fig:EF} (a) and (b) compare the experimental data on twinned hole peaks and the theoretically expected Landau spectrum in the two cases. As seen in Fig.\ref{fig:EF} (b),  assuming that the two twinned crystals share a common chemical potential set by the main crystal, the theoretical expectation is incompatible with the experimental results, in sharp contrast with the agreement seen between theory and experiment in the former case, see Fig.\ref{fig:EF}(a). 

This conclusion implies a discontinuity in chemical potential and in carrier density across the twin boundary, which is reminiscent of a Schottky barrier. Remarkably, when the magnetic field is along the trigonal axis, the difference in chemical potentials increases up to 68 meV at 65 T (see Fig.\ref{fig:EF} (c)). This is remarkable given that the twin boundary width is as narrow as 1 nm \cite{Edelman2008,Edelman2005}. Below, we turn to the underlying factors that impedes electrons belonging to adjacent crystals from mixing up.\\

\textbf{DISCUSSION}

We now propose a possible scenario explaining this effect. It is well known that the bands are bent near the interface between electronically different solids \cite{Kittel_book}. The typical examples are shown in Fig. \ref{Fig_boundary} (a) for the p-n junction and (b) for the Schottky junction. The former is a junction between semiconductors, and the latter is a junction between a metal and a semiconductor. In each case, chemical potentials are different in each side when separated (upper panels of Fig. \ref{Fig_boundary}). When they are joined, the carriers are forced to move to make $\mu$ common. For example, holes move from p-type to n-type, and electrons move from n-type to p-type in the p-n junction. The n-type band is shifted upward and the p-type band downward by the potential gradient due to the carrier transfer across the interface, resulting in the band bending with a common $\mu$ as shown in Fig. \ref{Fig_boundary} (a).

A similar band bending is also expected in the present twin boundary between Bi crystals. At zero field, $\mu$ is shared between the main and twinned crystals. Under a strong magnetic field, $\mu$ of the main crystal shifts differently from that of the twinned one, as shown in Fig. \ref{fig:EF} (c). This situation can be expressed by the model where two electronically different semimetals are joined at the twin boundary as depicted in Fig. \ref{Fig_boundary} (c).

Suppose that $\mu$ of the main crystal is lower than that of the twinned one when two semimetals are separated. When they are joined, the carriers are forced to move to make $\mu$ common. As a result, both the electron and hole bands are bent, as if two types of p-n junction coexist.
By this band bending, the relative levels between $\mu$, the conduction and the valence bands are kept in each crystal as before joining.
The Landau level spectrum of this twinned p-n junction is equivalent to that of the case where the independent chemical potentials are employed. Eventually, the mystery of the apparent jump of $\mu$ is solved by this scenario of the twinned p-n junction.

For the twinned p-n junction scenario, we assumed an electric potential difference between the main and twinned crystals. One may question whether an inhomogeneous electric field can exist inside a (semi) metal. The standard answer would be ``No" because finite current will screen the potential difference. However, Landauer argued against this simplistic view \cite{Landauer1975,Landauer1988}. Even in a metal, a point-like scattering center inside a metal will create an unscreened dipolar field, i.e., the electric field is inhomogeneous near the point-like defects. In the present case, the twin boundary, a line or plane defect, will play a similar role to the point-like scattering center of Landauer's theory.

In this scenario, the Landau eigenvalues curve in the vicinity of a twin boundary in the same way as they bend near the edges of a 2D electron gas \cite{MacDonald1983}. Any time of such Landau level bending results in edge currents, the continuity equation causes the edge charge and currents to remain close to the edge but distributed in real space in a 2D case\cite{MacDonald1983}. In the present 3D case, the continuity equation strictly confines the currents and any associated charge variations to the vicinity of the twin boundary. The currents flow along the twin boundaries and essentially account for small differences in magnetization between twins, which should be present due to differences in Landau level filling factors related to the difference in chemical potential. The equilibrium distribution of currents and charge ultimately depends on a balance between the Lorentz force of the current and the twin boundary potential that causes the Landau level eigenstates to curve in real space. These currents flowing along the twin boundary result from a form of extended bound state associated with the twin junction. This line of inquiry deserves further investigations.

The twinned p-n junction bears a resemblance to the p-n junction in semiconductors. Therefore, rectification effects observed in semiconductor p-n junctions are also expected in twinned p-n junctions. If these rectification effects are observed, it would provide direct evidence of the potential barrier we propose. However, in the present twinned p-n junction, significant rectification effects are not anticipated to be observed because they are semimetals, not semiconductors. This could explain why direct evidence of the twinned p-n junctions has not been obtained thus far.
The rectification effects may become observable if one type of carrier is evaporated, e.g., by applying a gate voltage.

The presence of twins in Bi has already been observed using scanning tunneling microscopy \cite{Edelman1996}. Their presence could also be detected using scanning photocurrent microscopy in presence of a magnetic field, though, to our knowledge, there has been no report of such observation yet. In either case, the presence of the potential barrier we propose may be challenging to detect when observing in the semimetallic state. For instance, applying a gate voltage and subsequently scanning after modifying electron-hole compensation in one domain might make detecting the potential barrier possible.

In summary, we have carried out angle-dependence of magnetoresistance along three high symmetric planes, and mapped out their Landau spectra up to 65 T. All anomalies including the one observed at $\sim$40 T for a magnetic field oriented along the trigonal axis could be explained by assuming the presence of a second twinned crystal and the two crystals keep separate chemical potentials. We suggested an interpretation of this result by a scenario in which a p-n junction builds up at the twin boundary-areas. This implies the existence of an inhomogeneous electric field in the vicinity of twin boundaries, which can be checked by future experiments.
\\
\\
\textbf{METHODS}\\
\textbf{Samples and measurements description}

Bismuth single crystals commercially obtained used in this study had typical dimensions of $1\times2\times0.5 \rm{mm}^3$ and a Room-temperature-to-Residual-resistivity-Ratio (RRR) of $\rho(300 {\rm{~K}})/\rho(4.2 {\rm{~K}}))\sim30-100$. Magnetoresistance (MR) measurements in pulsed magnetic field were carried out at Wuhan National High Magnetic Field Center (WHMFC) and in National High Magnetic Field Laboratory- Pulsed Field Facility (NHMFL-PFF) in Los Alamos. The samples were rotated in three distinct planes: binary-trigonal, bisectrix-trigonal and binary-bisectrix. For each set of measurements, the current was applied along the axis perpendicular to the rotating plane. Our high-field MR data complemented the Nernst data obtained below 28 T and reported previously \cite{Zhu2012a,Zhu2011}. Additional angle-dependent Nernst measurements up to 45 T were also carried out in the hybrid magnet of the NHMFL DC Facility in Tallahassee. Ultrasound measurements have been collected at Laboratoire National des Champs Magnétique Intense (LNCMI-Toulouse) with pulsed magnetic field up to 54 T. Longitudinal ultrasonic waves were generated using commercial LiNbO$_3$ 36$^\circ$ Y-cut transducers glued on a polished surface. A standard pulse-echo technique was used to determine the change in the sound velocity and attenuation. Magneto-optical measurements involving polarized light (for more details, see reference \cite{2010Katakura_MO}) were conducted at the International MegaGauss Science Laboratory, the Institute for Solid State Physics at the University of Tokyo. The resulting image was captured using a high-speed camera.

%\noindent
%* \verb|zengwei.zhu@hust.edu.cn|\\

\textbf{DATA AVAILABILITY}\\
The data supporting the present work are available from the corresponding authors
upon request.\\

\textbf{ACKNOWLEDGEMENTS}\\
This work was supported by The National Key Research and Development Program of China (Grants No.2022YFA1403500 and 2016YFA0401704), the National Science Foundation of China (Grant No. 11574097 and No. 51861135104),  and by directors funding grant number 20120772 at LANL. N.H. and RMcD acknowledge support from the US-DOE BES 'Science of 100T' program. B. F. acknowledges support from Jeunes Equipes de l’Institut de Physique du Coll\'ege de France (JEIP). K. B. was supported by the Agence Nationale de la Recherche (ANR-19-CE30-0014- 04). Y. F. was supported by JSPS KAKENHI grants 23H04862, 23H00268, and 18KK0132. Work performed at National High Magnetic Field Laboratory was supported by National Science Foundation Cooperative Agreements DMR-0654118 and DMR-2128556 and the State of Florida. M. T. acknowledges support from the JSPS KAKENHI grant No. 23H04862. \\
\textbf{COMPETING INTERESTS}\\
The authors declare no competing interests.\\
\textbf{AUTHOR CONTRIBUTIONS}\\
Z.Z. and K.B. conceived this work; Y.Y., Z.Z., J.W., P.N., L.X., H.Z., N.H., and R.D.M. performed the magnetoresistance experiment. A.Y. and Y.F. preformed the theoretical calculation. Y. K and M. T preformed the magneto-optics measurements. A.V.S, A.A., M.S.N, B.F., and K.B. preformed the Nernst effect measurement up to 45 T. B.F., D.L., and C.P. preformed ultrasound measurement. Z.Z., B.F., Y.F., and K.B. wrote the manuscript with input from all authors.\\

\clearpage

\renewcommand{\thefigure}{S\arabic{figure}}
\renewcommand{\thetable}{S\arabic{table}}
\def\theequation{S\arabic{equation}}

\makeatletter
\def\@hangfrom@section#1#2#3{\@hangfrom{#1#2}#3}
\def\@hangfroms@section#1#2{#1#2}
\makeatother

\renewcommand{\thesection}{Supplementary Note \arabic{section}}
\renewcommand{\thetable}{Supplementary Table \arabic{table}}

\renewcommand{\theequation}{Supplementary Equation \arabic{equation}}
\renewcommand{\figurename}{}
\renewcommand{\thefigure}{Supplementary Figure \arabic{figure}}
\setcounter{section}{0}
\setcounter{figure}{0}
\setcounter{table}{0}
\setcounter{equation}{0}

%\usepackage[showframe,%Uncomment any one of the following lines to test
%%scale=0.7, marginratio={1:1, 2:3}, ignoreall,% default settings
%%text={7in,10in},centering,
%%margin=1.5in,
%%total={6.5in,8.75in}, top=1.2in, left=0.9in, includefoot,

%\makeatletter
%\AtBeginDocument{\let\LS@rot\@undefined}
%\makeatother
%\maketitle	

\clearpage
\onecolumngrid
\begin{center}
	{\large Supplemental Material for\\[2mm] \bf High-field immiscibility of electrons belonging to adjacent twinned bismuth crystals}\\[3mm]
	%\SetTracking{encoding=*}{0}\lsstyle
	Yuhao Ye$^{1}$, Akiyoshi Yamada$^{2,3}$, Yuto Kinoshita$^{3}$, Jinhua Wang$^{1}$, Pan Nie$^{1}$, Liangcai Xu$^{1}$, Huakun Zuo$^{1}$, Masashi Tokunaga$^{3}$, Neil Harrison$^{4}$, Ross D. McDonald$^{4}$, Alexey V. Suslov$^{5}$, Arzhang Ardavan$^{6}$, Moon-Sun Nam$^{6}$,  David LeBoeuf$^{7}$, Cyril Proust$^{7}$,  Beno\^{\i}t Fauqu\'{e}$^{8}$, Yuki Fuseya$^{2}$, Zengwei Zhu$^{1,*}$ and Kamran Behnia$^{9}$ \\[2mm]
	%\SetTracking{encoding=*}{0}\lsstyle
	{\small {\it (1) Wuhan National High Magnetic Field Center and School of Physics, Huazhong University of Science and Technology, Wuhan 430074, China\\
			(2) Department of Engineering Science, University of Electro-Communications, Chofu, Tokyo 182-8585, Japan\\
			(3)Institute for Solid State Physics, The University of Tokyo, Kashiwa, Chiba 277-8581, Japan\\
			(4) MS-E536, NHMFL, Los Alamos National Laboratory,Los Alamos, New Mexico 87545, USA\\
			(5)National High Magnetic Field Laboratory, 1800 E. Paul Dirac Drive, Tallahassee, FL 32310, USA\\
			(6) Department of Physics, Clarendon Laboratory, University of Oxford, Oxford OX1 3PU, United Kingdom\\
			(7) Laboratoire National des Champs Magn\'{e}tiques Intenses (LNCMI-EMFL), CNRS, UGA, UPS, INSA,
			Grenoble/Toulouse, France\\
			(8) JEIP,  USR 3573 CNRS, Coll\`{e}ge de France, PSL Research University, 11, place Marcelin Berthelot, 75231 Paris Cedex 05, France\\
			(9) Laboratoire Physique et Etude de Mat\'{e}riaux (CNRS-UPMC)\\
			ESPCI Paris, PSL Research University, 75005 Paris, France}\\[0mm]}
\end{center}
\setcounter{page}{1}
\vspace*{5mm}

% Add 'S' to the numbering inside the supplement

\maketitle

%	\linenumbers
%	\pagewiselinenumbers
\twocolumngrid
\renewcommand{\thefigure}{S\arabic{figure}}
\renewcommand{\thetable}{S\arabic{table}}
\def\theequation{S\arabic{equation}}

\makeatletter
\def\@hangfrom@section#1#2#3{\@hangfrom{#1#2}#3}
\def\@hangfroms@section#1#2{#1#2}
\makeatother

\renewcommand{\thesection}{S\arabic{section}}
\renewcommand{\thetable}{S\arabic{table}}
\renewcommand{\thefigure}{S\arabic{figure}}
\renewcommand{\theequation}{S\arabic{equation}}

\setcounter{section}{0}
\setcounter{figure}{0}
\setcounter{table}{0}
\setcounter{equation}{0}
\maketitle

	\maketitle	
	%	\linenumbers
	%	\pagewiselinenumbers
	\section*{SUPPLEMENTARY NOTES}
	
	\subsection*{ Supplementary Note 1. Magneto-optical measurement on another sample}
	
\begin{figure}[ht]
		\setlength {\abovecaptionskip} {-0.2cm}
		\setlength {\belowcaptionskip} {-0.4cm}
		\includegraphics[width=9cm]{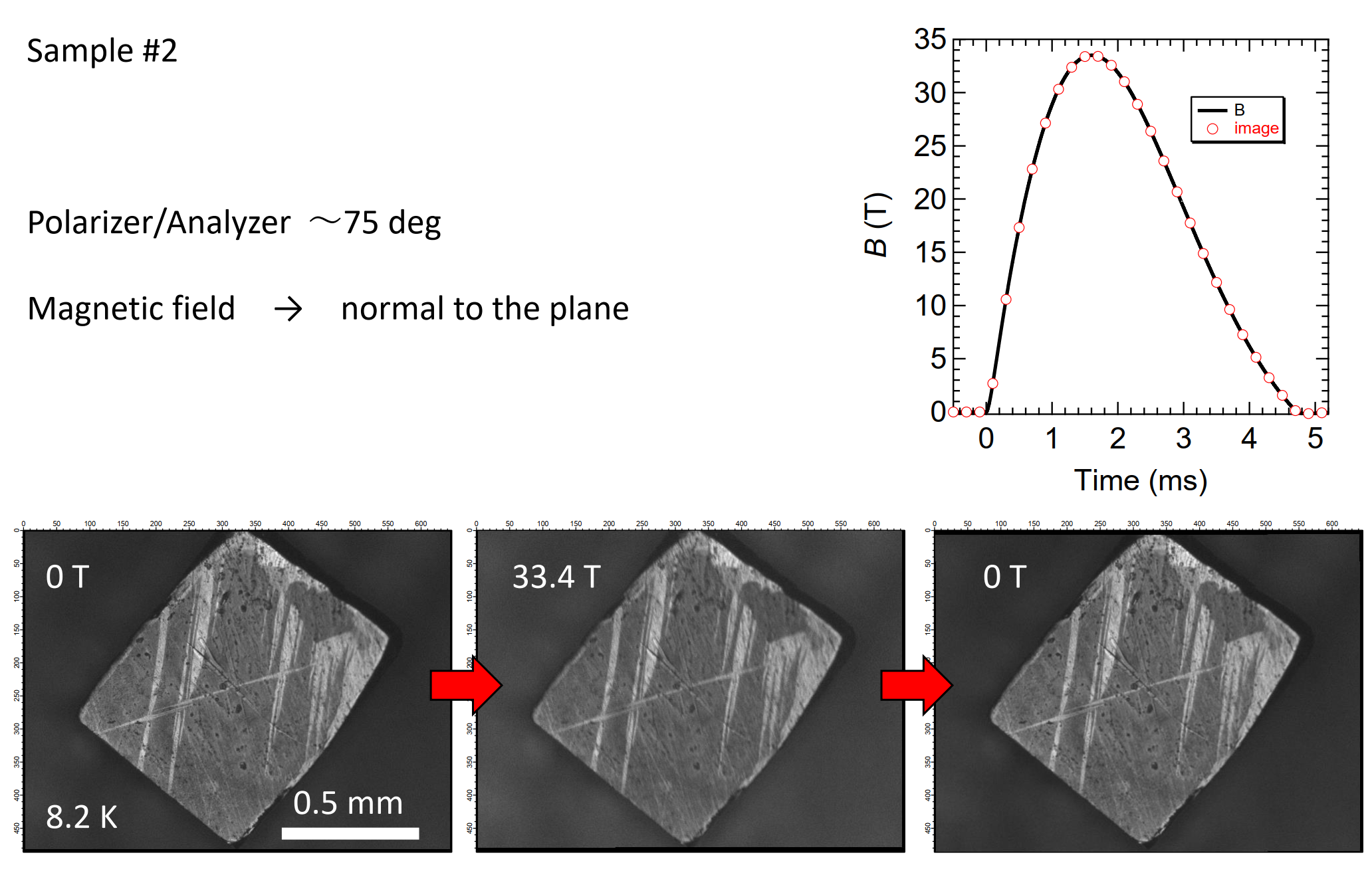}
		\caption{\textbf{No detectable change of twins under pulsed magnetic field in another sample named \#2 with magneto-optical imaging:} }
		\label{fig:S1}
	\end{figure}

	We measured additional magneto-optical results on another bismuth sample with same configuration as descried in the main text in the \ref{fig:S1}. The results are similar to those of the sample shown in the main text. 
	
	\subsection*{ Supplementary Note 2. The experimental and theoretical Landau spectrum in linear Y-scale}
\begin{figure*}[ht]
		\setlength {\abovecaptionskip} {-0.2cm}
		\setlength {\belowcaptionskip} {-0.4cm}
		\includegraphics[width=16cm]{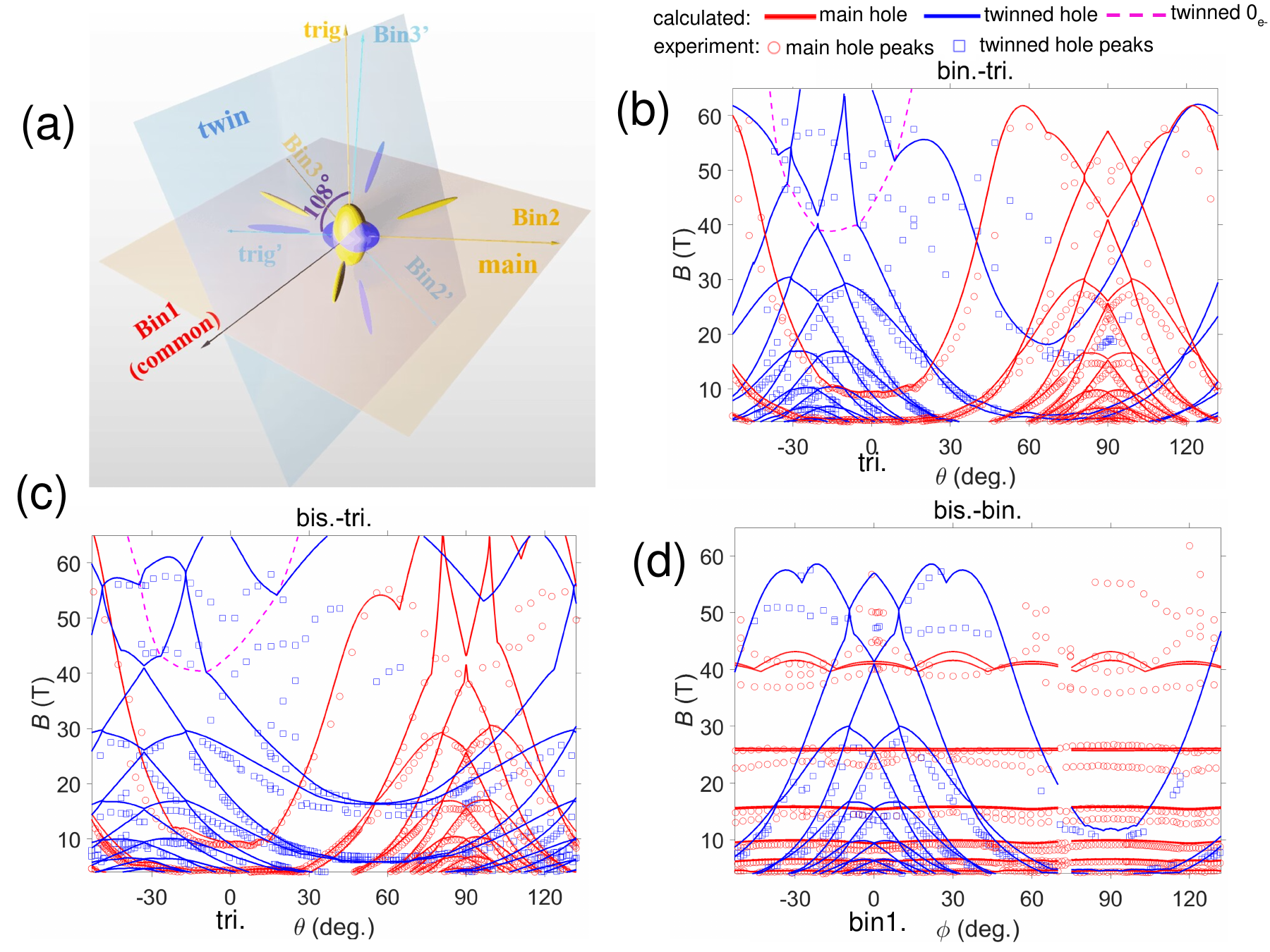}
		\caption{\textbf{The experimental and theoretical Landau spectrum: } (a) The illustration of the twinned structure in bismuth. The two crystals tilt from each other by 108$^{\circ}$ and share one common binary. (b) The Landau spectrum of the main and twinned samples from holes according to theory (solid lines) and experiment (symbols) for binary-trigonal, (c) bisectrix-trigonal, and (d) binary-bisectrix planes. The data are the same as the main text, but in linear scale for Y-axis. The red and blue symbols are for the peaks from the holes of the main and twinned crystals from the Fig. 3 of the main text, respectively. The red and blue lines represent the calculated Landau levels from holes of the main and twinned domains, respectively. The magenta dashed lines are for calculated 0$_{e-}$ electron Landau level which corresponds to the large drop in magnetoresistance as the field is along trigonal. The other electron peaks below 28 T are omitted for clarity, see the ref \cite{Zhu2012a}.}
		\label{fig:S2}
	\end{figure*}

	The \ref{fig:S2} shows the experimental and theoretical Landau spectrum is in linear Y-scale for magnetic field. The data are same as the Fig.2 of the main text. Four twinned domains are possible. However, our data can be fit by assuming the presence of a single minority domain in addition to the main crystal. It is possible that there are additional minority domains, but their population is below our detection level. It is also possible that during crystal growth pressure inhomogeneity favors the emergence of a minority domain along a specific orientation instead of three.

	\subsection*{Supplementary Note 3. The experimental and theoretical Landau spectrum including low-field electron spectrum}
\begin{figure*}[ht]
		\setlength {\abovecaptionskip} {-0.2cm}
		\setlength {\belowcaptionskip} {-0.4cm}
		\includegraphics[width=16cm]{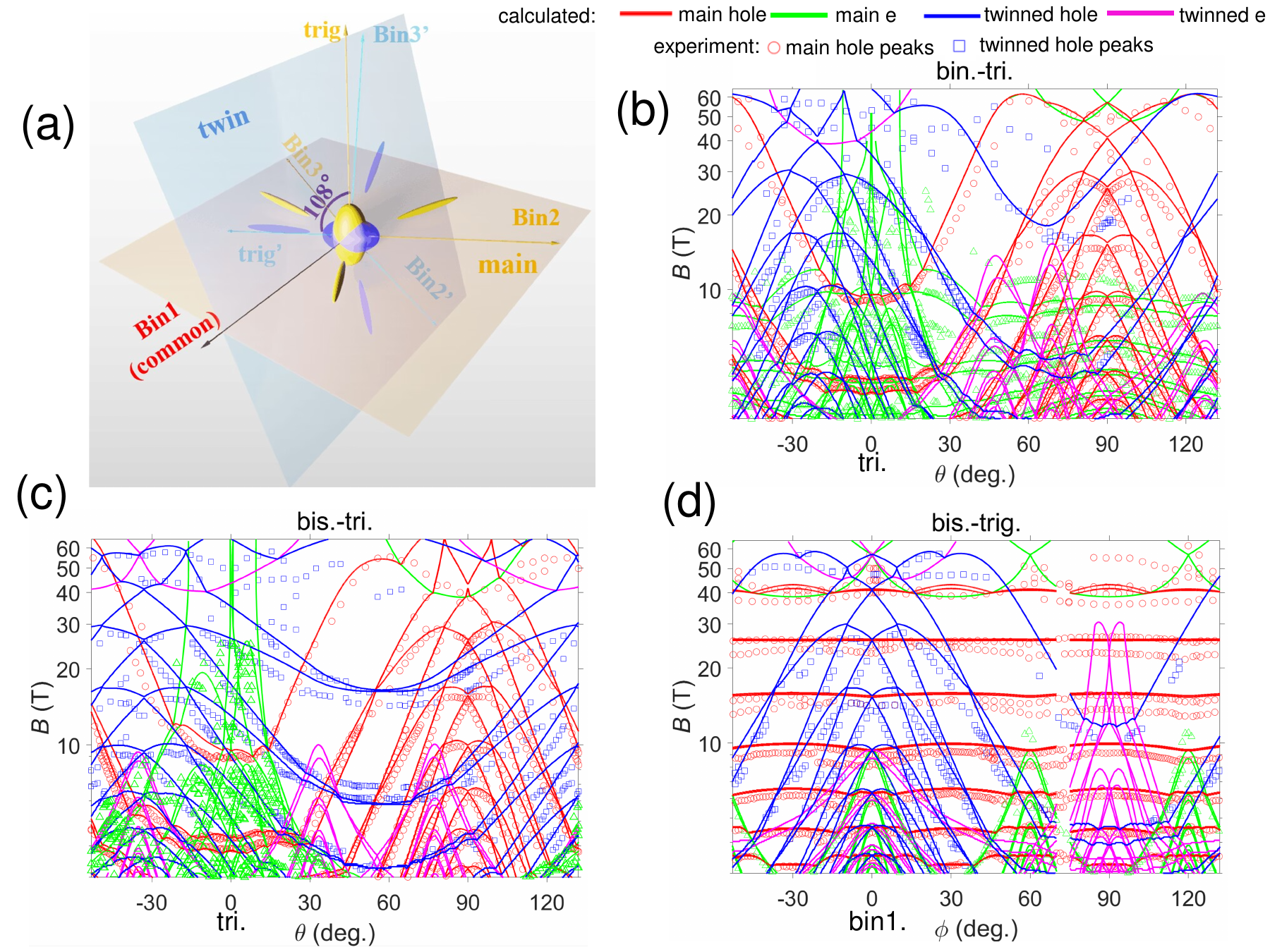}
		\caption{\textbf{The experimental and theoretical Landau spectrum: } (a) The illustration of the twinned structure in bismuth. The two crystals tilt from each other by 108$^{\circ}$ and share one common binary. (b) The Landau spectrum of the main and twinned samples from holes according to theory (solid lines) and experiment (symbols) for binary-trigonal, (c) bisectrix-trigonal, and (d) binary-bisectrix planes. The data are the same as the main text, but in linear scale for Y-axis. The red and blue symbols are for the peaks from the holes of the main and twinned crystals from the Fig. 3 of the main text, respectively. The red and blue lines represent the calculated Landau levels from holes of the main and twinned domains, respectively. The magenta dashed lines are for calculated 0$_{e-}$ electron Landau level which corresponds to the large drop in magnetoresistance as the field is along trigonal. }
		\label{fig:S3}
	\end{figure*}

	The \ref{fig:S3} shows the experimental and theoretical Landau spectrum including low-field electron spectrum. The complex of the Landau spectrum is further complicated by the electron spectrum. In high fields, the electron spectrum is hard to be distinguished from others. And a misalignment could partially explain the discrepancy seen in panels b and c. We also note that experiment cannot distinguish the origin (electron or hole, main or minority crystal) of an observed anomaly. Finally, since these anomalies are peaks in second derivative of magnetoresistance, some of them may be noise rather than signal. The agreement at high field is not as good as low field, which may be related to the higher sensitivity of the Landau spectrum to fine tuning at high field. From the comparison of theory and experiment for common and independent chemical potentials in the Fig.6 in the main text, it is safe to conclude that theory gives a qualitatively better account of the data in one case compared to the other.
	
	\begin{figure*}[ht]
		\setlength {\abovecaptionskip} {-0.2cm}
		\setlength {\belowcaptionskip} {-0.4cm}
		\includegraphics[width=17cm]{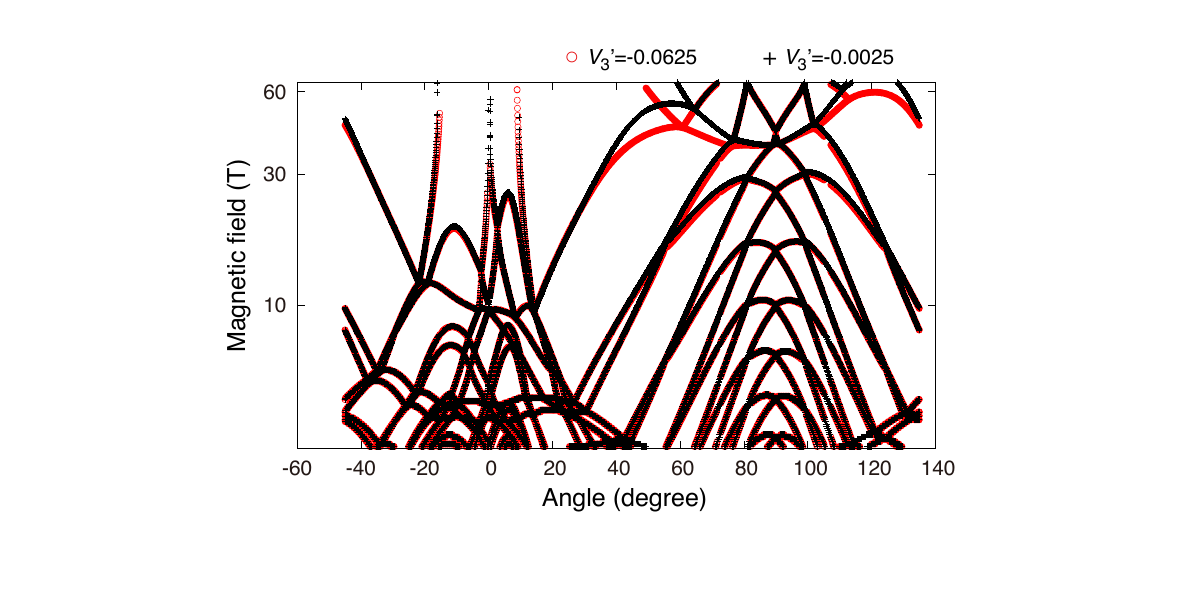}
		\caption{\textbf{Landau spectrum for main crystal in bis-tri plane with different $V_3$'.} Red and black points show the spectrum with $V_3$’= -0.0625 and -0.0025 respectively }
		\label{fig:S4}
	\end{figure*}
	\subsection*{ Supplementary Note 4. Different $V_3'$ for theoretical Landau spectrum}
	
	The \ref{fig:S4} shows the Landau spectrum for main crystal in bis.-tri. plane with different $V_3'$. Setting $V_3'$ to -0.0625, we could achieve a better fit to the experimental values around 40 Tesla. This parameter affects the lowest Landau level only and has no impact on other levels. Consequently, as shown in the figure, the spectra at low magnetic fields remain unchanged from the original parameters, and only a small part of spectra at the quantum limit is modified.
	Since the value of the g-factor is estimated from spectra below 30 Tesla, there is no change in the g-factor value with the adjustments made in this study.

\end{document}